\documentclass[journal]{vgtc}                     

\onlineid{0}

\vgtccategory{Research}

\title{Transition Techniques for \\Externally-Guided Multi-Scale Viewpoint Changes}

\author{%
  \authororcid{Matt Gottsacker}{0000-0002-3575-1133},
  \authororcid{Mengyu Chen}{0000-0001-6833-7273},
  \authororcid{David Saffo}{0000-0001-9515-048X},
  \authororcid{Feiyu Lu}{0000-0002-1939-9352}, 
  \authororcid{Benjamin Lee}{0000-0002-1171-4741}, and
  \authororcid{Blair MacIntyre}{0000-0002-5357-2366}
}
\authorfooter{
  \item All authors were with JPMorganChase when this work was done.
  \item
    MG is now with the University of Central Florida. Email: mattg@ucf.edu
  \item BM is with Northeastern University. Email:  b.macintyre@northeastern.edu
  \item FL current email: feiyulu@vt.edu
  \item  
  	All other authors are still with JPMorganChase.
  	E-mail: 
   \{mengyu.chen\,$|$\,david.saffo\,$|$\,benjamin.lee\}@jpmchase.com
}

\abstract{%
Extended reality (XR) is increasingly used to help users understand complex virtual environments through multiple viewpoints across different immersion levels, positions, and scales. While numerous techniques address viewpoint transitions for self-guided exploration, many scenarios require externally-guided transitions where a system or presenter controls the user's viewpoint, leaving the user with limited spatial knowledge and control over the transition process, which can increase susceptibility to disorientation and discomfort. We present three transition techniques for externally-guided multi-scale XR viewpoint changes and evaluate them against a fade-to-black baseline in a within-subjects study (N=20). Participants transitioned between world-in-miniature, street-level, and indoor destination views. We combined spatial recall measures, standardized questionnaires, and semi-structured interviews to assess orientation, workload, comfort, and continuity. Results showed that in our setup, techniques externalizing reference frames and the user's pose improved multi-scale spatial recall relative to a fade, while same-scale recall was insensitive to transition technique. We conclude with design implications for multi-scale XR viewpoint transitions.
}


\teaser{
  \centering
  \includegraphics[width=0.85\linewidth, alt={Schematic illustration of the study progression as three hand-drawn isometric scenes connected by arrows, left to right. In the first scene, labeled WiM View, a person wearing an XR headset looks down at a miniature city model floating in front of them. An arrow labeled Multi-Scale Transition points to the second scene, accompanied by three small icons of the transition techniques. Icon a shows the miniature map rotating in place in front of the stationary person. Icon b shows a translucent copy of the person walking away from their body, connected by a dashed line. Icon c shows a small flying camera tracing a dashed arc above a floating screen panel that the person watches. In the second scene, labeled Street View, the person stands at full scale on a city street looking up at a wireframe building with one room highlighted in solid blue. A second arrow, labeled Same-Scale Transition, points to the third scene, labeled Inside View, where the person stands inside a furnished office room with a desk, plants, and windows overlooking the city.}]{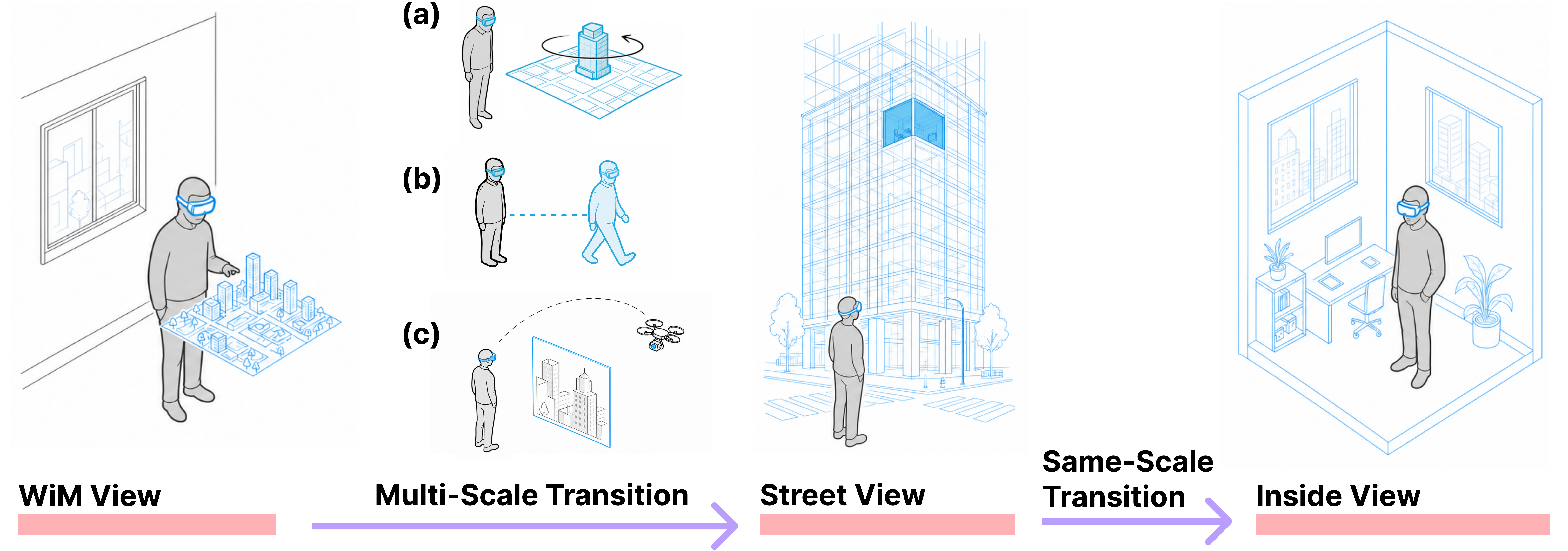}
  \cvspace{-1ex}
  \caption{%
  Users are guided through three viewpoints of a virtual environment. They begin with a world-in-miniature overview (\wimview{}), travel to a full-scale \streetview{} in front of a building, and finish inside a room of that building (\insideview{}). 
  We designed three techniques to keep users oriented and comfortable during these externally-guided transitions, each of which is used for both the multi-scale and same-scale transitions.
  (a)~\tObj{} rotates and translates the miniature environment in front of the stationary user to align it with the target viewpoint. (b)~\tAvatar{} sends an avatar out from the user's body to preview the target pose. (c)~\tFlycam{} shows a live feed of a virtual camera that flies to the target viewpoint. (Images generated with gpt-image-2 model.)
  }
  \label{fig:teaser}
}

\graphicspath{{figs/}{figures/}{pictures/}{images/}{./}} 

\usepackage{booktabs}                  
\usepackage{lipsum}                    
\usepackage{mwe}                       
\usepackage{ccicons}                   

\usepackage{mathptmx}                  

\usepackage{multirow}
\usepackage{gensymb}
\usepackage{wrapfig}
\usepackage{ifthen}
\usepackage[table]{xcolor}
\usepackage{amsmath}
\usepackage{colortbl} 

\definecolor{transitionrow}{HTML}{FFF3CC}  
\definecolor{spatialtestrow}{HTML}{E8FFE8} 

\newboolean{showrevisions}
\setboolean{showrevisions}{false}
\newboolean{isdraft}
\setboolean{isdraft}{false}

\newboolean{usevspace}
\setboolean{usevspace}{true}
\newcommand{\cvspace}[1]{\ifthenelse{\boolean{usevspace}}{\vspace{#1}}{}}

\ifthenelse{\boolean{showrevisions}}{\definecolor{newcolor}{rgb}{0.15, 0.15, 1}
\definecolor{chcolor}{rgb}{0.15, 0.5, 0.15}}{\definecolor{newcolor}{rgb}{0, 0, 0}
\definecolor{chcolor}{rgb}{0, 0, 0}}

\ifthenelse{\boolean{isdraft}}{\newcommand{\note}[1]{\textcolor{red}{\textbf{[#1]}}}}{\newcommand{\note}[1]{}}
\ifthenelse{\boolean{isdraft}}{\newcommand{\comment}[1]{\textcolor{blue}{\textbf{[#1]}}}}{\newcommand{\comment}[1]{}}
\ifthenelse{\boolean{isdraft}}{\newcommand{\new}[1]{\textcolor{blue}{#1}}}{\newcommand{\new}[1]{#1}}
\definecolor{freshcolor}{rgb}{0.0, 0.55, 0.0}
\ifthenelse{\boolean{isdraft}}{}{}

\newcommand{\secref}[1]{\hyperref[#1]{Sec.~\ref*{#1}}}
\newcommand{\figref}[1]{\hyperref[#1]{Fig.~\ref*{#1}}}
\newcommand{\tabref}[1]{\hyperref[#1]{Table~\ref*{#1}}}

\newcommand{\wimview}{\textsc{WiM View}}
\newcommand{\streetview}{\textsc{Street View}}
\newcommand{\insideview}{\textsc{Inside View}}

\newcommand{\tFade}{\textsc{Fade}}
\newcommand{\tFlycam}{\textsc{Flying Camera}}
\newcommand{\tObj}{\textsc{Object-Based}}
\newcommand{\tAvatar}{\textsc{Avatar}}

\begin{document}


\firstsection{Introduction}

\maketitle

Extended reality (XR) technologies such as augmented reality (AR) and virtual reality (VR) have been used to help users understand large and/or complex virtual environments (VEs) that require viewing structures from multiple vantage points, such as layouts of buildings~\cite{bruder2009archexplore, cheong_enhancing_2023}, 3D environments~\cite{chen_urbanrama_2021}, and spatial data~\cite{lee2021visceralization}.
These experiences often involve \new{virtually traveling between} different viewpoints of a VE, \new{at times across} different immersion levels, positions, and scales.

Such viewpoint transitions can occur in \textit{augmented presentations}, in which a knowledgeable presenter guides audience member(s) through XR content~\cite{gottsacker_examining_2025, vo_enhancing_2025}. These often use narrative visualizations with intentional structure and ordering for presenting key narrative details to the viewer~\cite{segel_narrative_2010}.
\new{For example, consider a presenter walking an audience through a proposed urban development (\figref{fig:teaser}). To begin,} the presenter may onboard the audience in AR to provide real-world interaction with the presenter and an introductory view of \new{the development VE as a world-in-miniature (WiM)~\cite{stoakley_virtual_1995}} before switching their \textit{immersion level} to VR for a fully-engaged view of the VE~\cite{pointecker_bridging_2022}.
Additionally, the audience can be guided to different virtual \textit{positions} for improved spatial understanding of the VE, especially when animations provide continuous spatial orientation cues during travel~\cite{bhandari_teleportation_nodate, rahimi_scene_2020, lee_evaluating_2020, medeiros_effects_2016}.
The audience can also be guided between \textit{scales} to move from global overview to close-up detail \new{(e.g., from the WiM overview to a street-level view)}, which supports a more holistic understanding of spatial content~\cite{zielasko2019desk, lee2021visceralization}.

\new{Each transition presents user interface challenges.}
Abrupt immersion changes can be jarring, as users must reorient to substantially different surroundings~\cite{von_willich_qualitative_2025}.
Virtual travel can induce discomfort, sickness, and disorientation~\cite{teixeira_investigating_2024}.
Moreover, when moving between scales, it can be difficult to maintain a coherent mental map of the entire scene~\cite{lee_designing_2023}.

Numerous \new{locomotion and transition} techniques have been proposed to address these issues for \textit{self-guided} viewpoint transitions in which the user has the freedom to explore XR environments.
In these cases, the user has some prior knowledge of their target spatial context, specifies the target viewpoint, and navigates by themselves.
However, many other cases \new{(such as the augmented presentation scenario above)} require the user's viewpoint to be \textit{externally-guided}.
\new{
We define \textbf{externally-guided viewpoint transitions} as those in which an external agent (e.g., a system, program, or another user) controls any component of the user's virtual travel as described by Bowman et al.'s taxonomy~\cite{bowman1997travel, laviola20173d}, i.e., its initiation (input conditions), velocity/acceleration, and/or direction/target.
This transition class includes cases studied in related work, such as system-automated transitions~\cite{rahimi_scene_2020} or group navigation~\cite{weissker2020getting}.
In practice, such transitions appear in settings like financial data presentations~\cite{gottsacker2023presentation} and large-audience XR experiences that require consistent pacing for everyone~\cite{batra_xrxl_2025}.}

\new{The viewpoint transition issues described above are especially important in these scenarios.} The user has \textbf{limited control and spatial knowledge} of the target context and transition dynamics (e.g., how/when they transition).
They are thus less able to anticipate or predict their target viewpoint and path, and are more susceptible to negative user experience outcomes such as discomfort and disorientation~\cite{teixeira_investigating_2024}.
%
%
A key challenge with this setup is visualizing the target spatial context during the transition to support continuous understanding and avoid a disorienting experience that may distract from the presentation content.
Considering these user experience factors, our driving research question is:
\textbf{how should externally-guided viewpoint changes be performed so that the user remains oriented and comfortable while transitioning through changes in immersion, position, and scale?}

To address this question, we created three transition techniques designed to support a good user experience during externally-guided viewpoint transitions: \tObj{}, \tAvatar{}, \tFlycam{}; as well as a baseline \tFade{}.
We investigated their effects on users' spatial recall, workload, comfort, and perceived continuity when transitioning between WiM, street-level, and indoor destination views.
We report a controlled within-subjects study comparing our transition techniques, combining objective spatial recall measures, standardized questionnaires capturing user experience, and qualitative post-study interviews to explain how particular transition cues support orientation and a smooth user experience.
Our results show that transitions that externalize reference frames and the user's pose (\tObj{}, \tAvatar{}) can improve multi-scale spatial recall relative to a minimal fade, while same-scale recall was largely insensitive to transition type \new{in the arrangement we tested}.
We conclude with design implications for transition techniques that maintain continuity and reduce the cognitive work of reorientation during externally-guided multi-scale viewpoint changes in immersive environments.

In summary, our contributions are:
\begin{itemize}
    \item Designs and prototypes of externally-guided multi-scale transition techniques informed by related research.
    \item Empirical evaluation of four transition techniques in an \new{externally-guided} immersive presentation scenario with objective spatial recall metrics, subjective questionnaires, and 
    interviews.
    \item Design implications for building transition techniques that maintain continuity, support orientation, and minimize discomfort during \new{externally-guided} multi-scale viewpoint changes.
\end{itemize}

\section{Background}
\label{sec:background}

Related literature on transitions in viewpoint immersion level, position, and scale reveals several user experience considerations relevant to externally-guided contexts.

\subsection{Transitions Between Immersion Levels}

To enable the benefits of transitioning between different levels of immersion~\cite{pointecker_bridging_2022} while balancing the potential drawbacks like disorientation~\cite{von_willich_qualitative_2025}, researchers have explored numerous transitional interfaces.
For instance, foundational work showed transitions between VEs can establish and maintain presence~\cite{slater1994depth}.
Additionally, the MagicBook~\cite{billinghurst2001magicbook} was an early system that enabled a user to transition between viewing content in both AR and VR for different perspectives on a narrative.

Subsequent work has empirically validated these ideas and identified key usability insights.
For example, Steinicke et al. found that gradual transitions from the real world (RW) into VR can increase presence~\cite{steinicke2009gradual} and distance judgments~\cite{steinicke2009transitional}.
Similarly, the concept of Smooth Immersion from Valkov et al.~\cite{valkov2017smooth} argues for smooth, continuous RW--VR transitions as a mechanism for improving the overall experience of entering VEs.
A related line of work focuses on the process of exiting VR, which has been characterized as potentially jarring and multidimensional~\cite{knibbe2018dream}, motivating transition techniques that ease exit from VR and re-entry to the physical world~\cite{horst2021back, piumsomboon2022excitxr}.

The efficiency of transitions also matters.
For example, related work has explored interaction techniques for quickly switching between different immersion levels~\cite{das_exploring_2024} and VEs~\cite{gottsacker_one_2026}. 
Researchers have also explored different transition effects in task-based contexts, finding that users often prefer simple, efficient fade-to-black transitions~\cite{pointecker_bridging_2022, feld2024simple}.

\subsection{Transitions Between Positions}

Traveling between different positions within a VE is a core XR interaction for which many techniques have been designed~\cite{laviola20173d}. 
Travel transitions that visualize viewpoint motion to the user introduce fundamental trade-offs between spatial cognition and comfort.
Teleportation, which instantaneously changes the user's viewpoint without generating optical flow, is the dominant locomotion technique in consumer VR due to its minimal cybersickness risk~\cite{al_zayer_virtual_2020}.
However, the lack of intermediate visual cues impairs spatial updating: users receive no information about the path between origin and destination, which degrades path integration~\cite{bhandari_teleportation_nodate} and spatial awareness of the surrounding scene~\cite{rahimi_scene_2020}.
An alternative approach is continuous camera motion, in which the virtual viewpoint moves along a path from origin to destination.
This preserves the continuous optic flow that supports spatial updating, allowing users to track their changing position relative to landmarks~\cite{rahimi_scene_2020}.
However, optic flow without corresponding vestibular input produces vection (the illusory perception of self-motion), which is a primary driver of cybersickness under the sensory conflict model~\cite{zhao_mitigation_2023, nie_peripheral_2025}.

Researchers have proposed hybrid techniques to mitigate sickness during continuous motion through reduced field-of-view (FOV)~\cite{fernandes2016sickness}, persistent rest frames~\cite{cao_visually-induced_2018}, or peripheral visual elements~\cite{zhao_mitigation_2023}.
A complementary family of techniques augment teleportation with spatial cues to recover orientation information. Griffin and Folmer~\cite{griffin_out--body_2019} introduced out-of-body locomotion, in which users navigate a visible avatar from a stationary third-person view. Huang et al.~\cite{huang_preview_2025} presented Preview Teleport, which displays a preview window of the target viewpoint.

The work described above helps to reduce cybersickness and improve spatial cognition during self-guided viewpoint transitions.
Rahimi et al.~\cite{rahimi_scene_2020} studied system-automated viewpoint transitions and found the same fundamental relationships as described above: that animated interpolation supported the best spatial awareness but caused the most sickness, while teleportation produced the least sickness but the worst spatial awareness.
In externally-guided contexts such as presentations, the user lacks anticipatory knowledge of the destination, path, and timing, which can increase discomfort and disorientation risk~\cite{teixeira_investigating_2024}.
It remains unclear what transition designs might preserve orientation and comfort in such scenarios.

\subsection{Transitions Between Scales}

While the techniques described above are designed for virtual travel within a VE, environments with substantial scale or complexity require additional techniques for efficient XR viewpoint transitions across different levels of detail (e.g., top-down view and full-scale immersive view) so users can relate local observations to global structure.
An early example is world-in-miniature (WiM), which enables users to view and interact with an entire VE at a smaller scale, including selecting and flying to different viewpoints~\cite{stoakley_virtual_1995}.

More recent work has focused on techniques for optimizing transitions between scales for spatial understanding and user experience. 
For example, Weissker et al.~\cite{weissker_try_2024} extended teleportation with scale selection and attached a WiM cutout preview to the non-dominant hand to communicate the user's future surroundings at the selected scale.
Lee and Stuerzlinger~\cite{lee_scaling_2025} designed scroll-based exocentric scale control that separates scale-level assignment from scaling-center placement, with an X-ray visualization for seeing inside nested structures.
Additionally, techniques such as GulliVR~\cite{krekhov_gullivr_2018} enlarge the user's virtual body and virtual eye distance on demand to produce a miniature world perception in which physical walking is mapped to large-distance travel.
Cmentowski et al.~\cite{cmentowski2019outstanding} extended this idea to dynamic perspective switching between first-person and a scaled third-person bird's eye view using a smooth curved transition with proportional eye-distance scaling. 
In addition, supplementary tools such as mini-maps, portal previews of target locations, and X-ray building translucency can also support orientation and feedforward during navigation in large, occluded environments~\cite{shahbaz_badr_empirical_2023}.

Most directly relevant to our work, Lee et al.~\cite{lee_designing_2023} presented the first systematic investigation of viewpoint transition design for multi-scale VEs with nested structures.
They identified and evaluated three design components: transition trajectory, interactive timing control, and adaptive speed modulation. They also introduced methods for efficiently navigating and zooming among different positions and scales with hierarchical relationships.
A key design principle from this work is that separating rotation and translation phases during scale transitions helps maintain spatial orientation and reduce disorientation, and that VE rotation should occur at larger scales.

The scale transition techniques reviewed above are designed for self-guided exploration in which the user controls timing, scale, and destination.
Key affordances such as interactive timing control~\cite{lee_designing_2023}, user-initiated scale specification~\cite{weissker_try_2024, lee_scaling_2025}, and preview of target surroundings~\cite{weissker_try_2024, shahbaz_badr_empirical_2023} are absent or constrained in externally-guided contexts where a system or presenter controls the viewpoint, motivating additional exploration into how to preserve spatial orientation across multi-scale transitions when the user lacks both path control and prior knowledge of the spatial context.

\begin{table*}[ht]
\centering
\caption{Mapping of transition techniques to design considerations. \new{Empirically grounded takeaways for each technique appear in \secref{sec:implications}.}}
\cvspace{-1ex}
\label{tab:technique-dc-mapping}
\small
\begin{tabular}{l p{2.75cm} p{3.75cm} p{3.5cm} p{3.5cm}}
\toprule
& \textbf{[DC-1] Orientation} & 
\textbf{[DC-2] Comfort} &
\textbf{[DC-3] Aesthetic Continuity} &  
\textbf{[DC-4] Feedforward} \\
\midrule
\tFade{} & 
    No spatial cues & 
    No visual motion & 
    No continuity & 
    No preview \\
\tObj{} & 
    Map externalizes rotation & 
    User stationary, map moves & 
    Continuous geometric transform & 
    Target position and building, map \\
\tAvatar{} & 
    Third-person pose preview & 
    User stationary, avatar moves & 
    Embodied interaction & 
    Target position, map, body pose  \\
\tFlycam{} & 
    Camera shows travel path & 
    Decoupled frame; reduced FOV; mono & 
    Uninterrupted visual flow & 
    Target position, map \\
\bottomrule
\end{tabular}
\cvspace{-2ex}
\end{table*}

\section{Viewpoint Transition Techniques}
\label{sec:techniques}

We designed viewpoint transitions for both multi-scale and same-scale changes. In our presentation scenario, users begin in the real world viewing a WiM of a VE in AR, transition into a full-scale VR street-level view, and then to an indoor destination within the VE (\figref{fig:teaser}).
We developed our techniques through iterative prototyping and pilot testing, guided by design considerations drawn from related work (\secref{sec:background}).

\subsection{Design Considerations}

Prior work highlights the importance of spatial cognition support during viewpoint transitions~\cite{rahimi_scene_2020, bhandari_teleportation_nodate}, so we aimed to provide transitions that help users continuously \textbf{maintain spatial orientation and awareness (DC-1)}.
To this end, we designed transitions that gradually introduce structural elements of the target environment, maintain stable reference frames when possible, and avoid presenting users with a sudden, fully unfamiliar scene.
Additionally, prior work shows that visual--vestibular mismatches during virtual travel can cause discomfort, disorientation, or simulator sickness~\cite{cao_visually-induced_2018, lin_natural_2002, teixeira_investigating_2024}.
Common approaches to cybersickness reduction such as FOV restrictors proved insufficient in pilot testing: users reported feeling out of control, likening the experience to a rollercoaster.
To \textbf{minimize discomfort (DC-2)}, we constrained transitions so that scene motion remains congruent with the user's own motion, i.e., limiting how much the scene content can shift relative to the user’s physical head and body motion.

Research in XR transitions emphasizes the importance of hedonic qualities such as smoothness~\cite{valkov2017smooth}, visual polish~\cite{pointecker_bridging_2022}, and perceived continuity~\cite{husung2019portals}.
Thus, we aimed to \textbf{maximize perceived aesthetic continuity (DC-3)} through transitions that feel pleasant, are free of visual clutter, and provide an overall sense of flow between viewpoints.
Additionally, since unexpected virtual motion can contribute to discomfort~\cite{teixeira_investigating_2024}, we sought to provide \textbf{feedforward cues (DC-4)} that preview the outcome of the transition to form expectations about the upcoming spatial shift~\cite{muresan_using_2023}.
Accordingly, we designed each transition to visualize the destination context before the user's viewpoint change, giving users time to mentally prepare even in externally-guided scenarios where the user does not initiate the travel themselves.

\begin{figure*}[t]
    \centering
    \includegraphics[width=\linewidth]{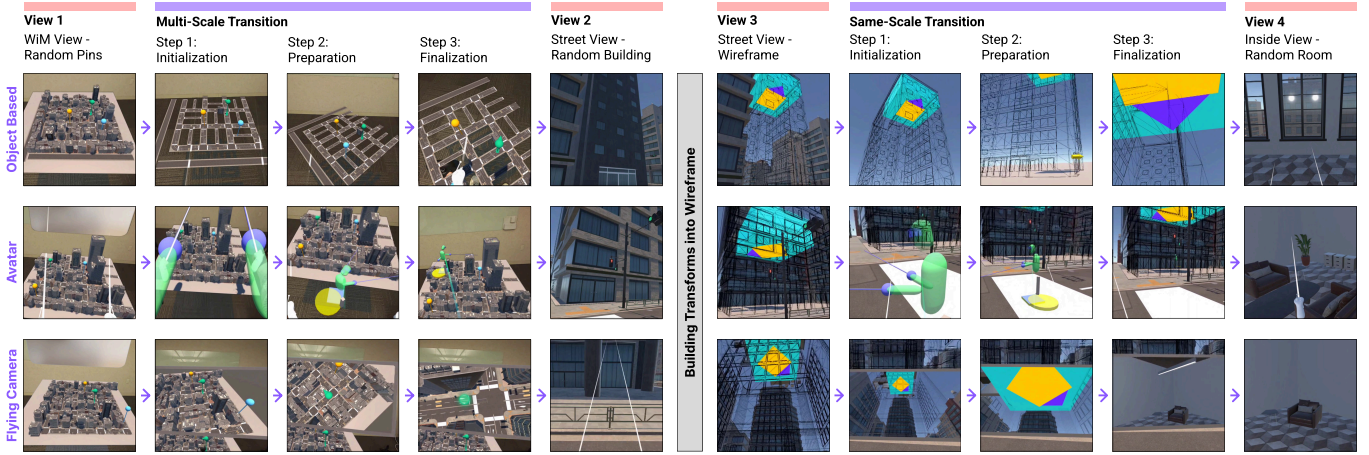}
    \caption{First-person view screenshots of a user experiencing each transition technique.
    }
    \label{fig:transition_flows}
    \cvspace{-1ex}
\end{figure*}

\subsection{Transition Approaches}

Our design considerations are not independent (e.g., a transition that provides rich spatial cues may also increase visual motion and thus risk discomfort) so any single technique must negotiate trade-offs among them.
Instead of optimizing for one concern in isolation, we explored three distinct strategies for resolving these trade-offs, each using a different mechanism for visualizing the spatial transition between the user's current and target viewpoints.
Here, we describe our approaches, supporting rationale, \new{positioning with respect to Bowman et al.'s virtual travel taxonomy~\cite{bowman1997travel, laviola20173d},} and implementation details (including pilot findings).
\new{\tabref{tab:technique-dc-mapping} summarizes how each technique maps to the design considerations.}
Technique transition stages are illustrated in~\figref{fig:transition_flows}, and technique videos are provided in the Supplementary Materials.


Each transition technique was divided into three steps.
The first step familiarized the user with the setup of the transition by introducing its main visual components.
Step~2 executed the primary animation, e.g., moving virtual objects to visualize the process of traveling to the target viewpoint.
Step~3 completed the transition with final animation motions (e.g., ascent or descent into the target viewpoint) and a fade into the target viewpoint.
\new{In virtual travel taxonomy terms~\cite{bowman1997travel}, the target viewpoint selection was always performed by the system and initially visualized through a position and direction cue, and the presenter triggered the start of the transition. 
The transitions differed in how the system controlled the velocity and acceleration of the virtual travel.}

In addition to our techniques, we included a \textbf{\tFade{} baseline} that fades to black and teleports the user to the target viewpoint \new{without visualizing virtual travel} and provides no spatial orientation, continuity, or feedforward cues.
This represents a minimal-information condition to compare against our spatially informative techniques.
It is similar to baseline transitions tested in related studies~\cite{gottsacker2024residue, feld2024simple}, which have been found to be most efficient and preferred in task-driven contexts~\cite{feld2024simple}.

\subsubsection{\tObj{}}
\textbf{In this approach, the scene moves around the stationary user rather than the user moving through the scene.}
The VE rotates and translates to align with the user's target heading before the viewpoint change occurs to externalize the mental rotation that would otherwise be required (DC-1).
The user remains physically and virtually stationary throughout to avoid visually-induced self-motion (DC-2).
The visuals of the VE components (buildings, etc.) are simplified to reduce clutter (DC-3) and emphasize the user's target position (DC-4).

\paragraph{\textbf{Rationale}} 
Heading-up or forward-up maps reduce mental rotation demands and improve directional decision-making because the depicted environment is already aligned with the user's facing direction, whereas north-up views require additional cognitive effort~\cite{montello2010where, smets2008mobilemap, levine1982youarehere, gagnon2014steppingintomap, Aretz1992mentalrotation}. 
Darken and Cevik~\cite{darken1999VEorientation} confirmed this advantage in VEs, finding that forward-up maps produced faster and more accurate targeted navigation than north-up maps by eliminating the mental rotation required to align egocentric and world reference frames.
Object-based transitions apply this principle to viewpoint changes: by rotating the VE to align with the user's eventual ground-level heading before entering the scene, the system externalizes the necessary mental rotation and preserves an egocentric reference frame across the view change.

\paragraph{\textbf{Implementation}}
For the multi-scale transition, some pilot users reported that translating the WiM under them led to minor discomfort due to feeling as though ``the rug was being pulled out'' from under them. 
We found that dramatically reducing the VE visuals helped preserve a stable reference frame.
Thus, during the multi-scale transition, every building except the one that the user would travel to was removed from the map. Other VE features (trees, signs, etc.) were also removed such that the user could only see the building they would travel to, the map streets, and the landmark pins.
For the same-scale transition, we achieved the visual effect of the VE rotating to align with the user's target viewpoint by rotating the user around the building.
For this, all VE features except for the wireframe building were removed to prevent the user from experiencing vection.
In pilots, users reported having a difficult time tracking how they had rotated relative to the building, so we added a yellow arrow that always pointed to their original \streetview{} position. \textbf{Steps:}

\begin{enumerate}[itemsep=0pt, parsep=0pt, topsep=0pt]
    \item VE visuals are reduced.
    \item VE rotates to align the target viewpoint's heading with the user's forward direction, then translates horizontally to position the target viewpoint above the user's head.
    \item VE moves vertically toward the user, indicating they are about to enter the target viewpoint.
\end{enumerate}
\noindent
\new{In travel taxonomy terms~\cite{bowman1997travel}, the system applied the travel's velocity/acceleration to the VE rather than to the user's viewpoint. It first animated the travel's rotational component by rotating the VE, then its positional component by translating the VE toward the user.}

\subsubsection{\tAvatar{}}
\textbf{This approach previews the user's future pose through an avatar representation.}
An abstract avatar departs the user's body, travels to the target location, and assumes the target orientation before the user is transitioned there.
The avatar's head and arms track with the user's motion, and its final pose gives a third-person preview of the destination position and heading, supporting orientation (DC-1).
Because the user's own viewpoint remains stationary while the avatar moves, the technique externalizes travel motion and avoids visual--vestibular conflict (DC-2).
The avatar's simple appearance and continuous motion provide visual flow across the transition (DC-3).
Users see where and how they will be positioned before the transition executes (DC-4).

\paragraph{\textbf{Rationale}}
Research on perspectives in VR has found that third-person viewpoints provide better global spatial awareness than first-person views, while first-person views support stronger embodiment and more efficient/precise interaction~\cite{gorisse_first-_2017, medeiros2018headshoulders,gottsacker_one_2026}. 
Multi-perspective travel techniques that let users command an avatar from a scaled third-person view and then re-embody it have reported spatial orientation benefits without harming presence or comfort~\cite{cmentowski2019outstanding, elvezio2017preoriented,gottsacker_one_2026}. 
Complementary work on self-avatars in AR and VR has shown that a visible self-avatar positioned in the environment can support body ownership and self-location~\cite{rosa_embodying_2019} and improve users' accuracy in judging spatial affordances at a distance~\cite{genay2022whatcanido}. 
The \tAvatar{} transition combines these ideas: an abstract, user-tracked avatar travels to the destination and assumes the target pose, giving users a third-person spatial preview of their future position and heading before the viewpoint changes.

\paragraph{\textbf{Implementation}}

When piloting, we found it was important to help communicate that the avatar was connected to the user and tracked their movements.
Thus, we included a gradual fade-in of the avatar materials and a line connecting the avatar's body to the user's. 
Unlike \tObj{}, this technique operates identically for both multi-scale and same-scale transitions.
\textbf{Steps:}

\begin{enumerate}[itemsep=0pt, parsep=0pt, topsep=0pt]
    \item The avatar fades in over the user's body.
    \item The avatar moves into the user's field of view and rotates to match the target viewpoint's orientation.
    The avatar then translates horizontally until vertically aligned with the target viewpoint.
    \item The avatar translates vertically to the target viewpoint.
\end{enumerate}
\noindent
\new{In travel taxonomy terms~\cite{bowman1997travel}, the system applied the travel's velocity/acceleration to the avatar proxy, which first rotated to the target orientation and then translated horizontally and vertically to the target viewpoint, while the user's viewpoint remained stationary.}

\subsubsection{\tFlycam{}}
\textbf{This approach shows the user's travel path through a decoupled virtual camera feed.}
A virtual camera flies along the travel path while the user watches its live feed on a flat, monoscopically rendered panel.
The camera path shows the spatial relationship connecting the two viewpoints (DC-1).
The decoupled reference frame, restricted field of view, and monoscopic rendering reduce discomfort risk by limiting visually-induced vection (DC-2).
The uninterrupted visual flow between scales maintains experiential continuity (DC-3).
The camera's live feed progressively reveals the destination context as the camera approaches it (DC-4).

\paragraph{\textbf{Rationale}}
Visually-induced vection without corresponding vestibular input is a primary driver of cybersickness~\cite{kelly_field_2025}, and the risk increases when the onset of motion is unexpected or poorly predicted~\cite{teixeira_investigating_2024}, which is a concern in externally-guided contexts where users cannot anticipate the onset, direction, or velocity of viewpoint motion. The \tFlycam{} technique addresses this by externalizing all travel motion to a world-referenced virtual camera: users observe the motion through a live feed on a flat panel instead of experiencing the motion from a first-person perspective. The panel's restricted field of view and monoscopic rendering further limit vection by reducing peripheral optical flow and depth cues associated with self-motion~\cite{kelly_field_2025, luu_effects_2021}.

\paragraph{\textbf{Implementation}}

For the multi-scale transition, a copy of the VE is rendered to a separate layer so that the camera's VE can scale independently without affecting the stable scene in front of the user. To maintain path legibility, the camera follows a three-segment path: ascending, translating horizontally to the destination, and descending to street level. This avoids interpolating directly through occluding geometry. 
For the same-scale transition, the camera travels directly to the target position before rotating to the final orientation.
\textbf{Steps:}

\begin{enumerate}[itemsep=0pt, parsep=0pt, topsep=0pt]
    \item A flat panel appears showing a live camera feed of the user's current surroundings.
    \item The camera departs and flies along the path to the target viewpoint; its feed progressively reveals the destination.
    \item The user is teleported to the target viewpoint at the end of the camera’s movement.
\end{enumerate}
\noindent
\new{In travel taxonomy terms~\cite{bowman1997travel}, the system applied the travel's velocity/acceleration to the decoupled camera, which traversed the full path between the viewpoints. The scale change was animated by scaling the camera's copy of the VE, while the user's own viewpoint remained stationary until the final teleport.}

\section{User Study}
\label{sec:userstudy}

To investigate the effects of our transition techniques on users' experiences and spatial orientation abilities, we conducted a within-subjects experiment ($N=20$).
Each participant experienced four transition techniques in four different but similarly structured VEs.
The order of techniques was counterbalanced using a Latin square, and VE order was randomized per participant.
Each condition involved three trials, each consisting of visiting a randomly chosen point-of-interest (POI) and completing spatial orientation tests.
Trial steps are illustrated in~\figref{fig:teaser} and enumerated in~\tabref{tab:trial_procedure}. First, the participant viewed the VE as a WiM (\wimview{}), transitioned to a full-scale street-level view (\streetview{}) in front of a building, and then transitioned to inside a room in the building (\insideview{}).
An experimenter triggered each transition.
\new{This fixed overview-then-detail progression mirrors common presenter workflows~\cite{segel_narrative_2010} and ensured that the multi-scale and same-scale transitions occurred at consistent positions within each trial. Reverse or randomized progressions would have introduced order effects that are not the target of this work, so we scoped the study to this single progression (see \secref{sec:limitations}).}
All transitions for the three POIs within a given VE used the same technique.

\subsection{Participants}

Our institution's ethics committee approved our study.
We recruited 20 participants from our institution (12 male, 8 female) aged 20--60+, with a median age range of 30--39. Five participants were aged 20--29, six were 30--39, five were 40--49, three were 50--59, and one was 60 or older. Participants self-reported their ethnicity as White (50\%), Asian (30\%), Black or African American (10\%), Hispanic or Latino (5\%), and one participant (5\%) did not indicate ethnicity.

\subsection{Hardware and Setup}

We ran the experiment system using Unity version 2022.3.4 on a Dell Precision 7680 laptop equipped with an Intel i9 CPU, an NVIDIA RTX 4000 Ada Generation Laptop GPU, and 64 GB of RAM.
The laptop was connected to an HTC VIVE XR Elite HWD with a VIVE Streaming Cable.
This HWD supports both VR and full-color video see-through AR, has a resolution of $1920\times1920$ per eye, a refresh rate of 90 frames per second, a diagonal field of view of 110\degree, and accommodates interpupillary distance in the range of 54 to 73 mm.

\begin{figure}[t!]
    \centering
    \includegraphics[width=0.85\linewidth]{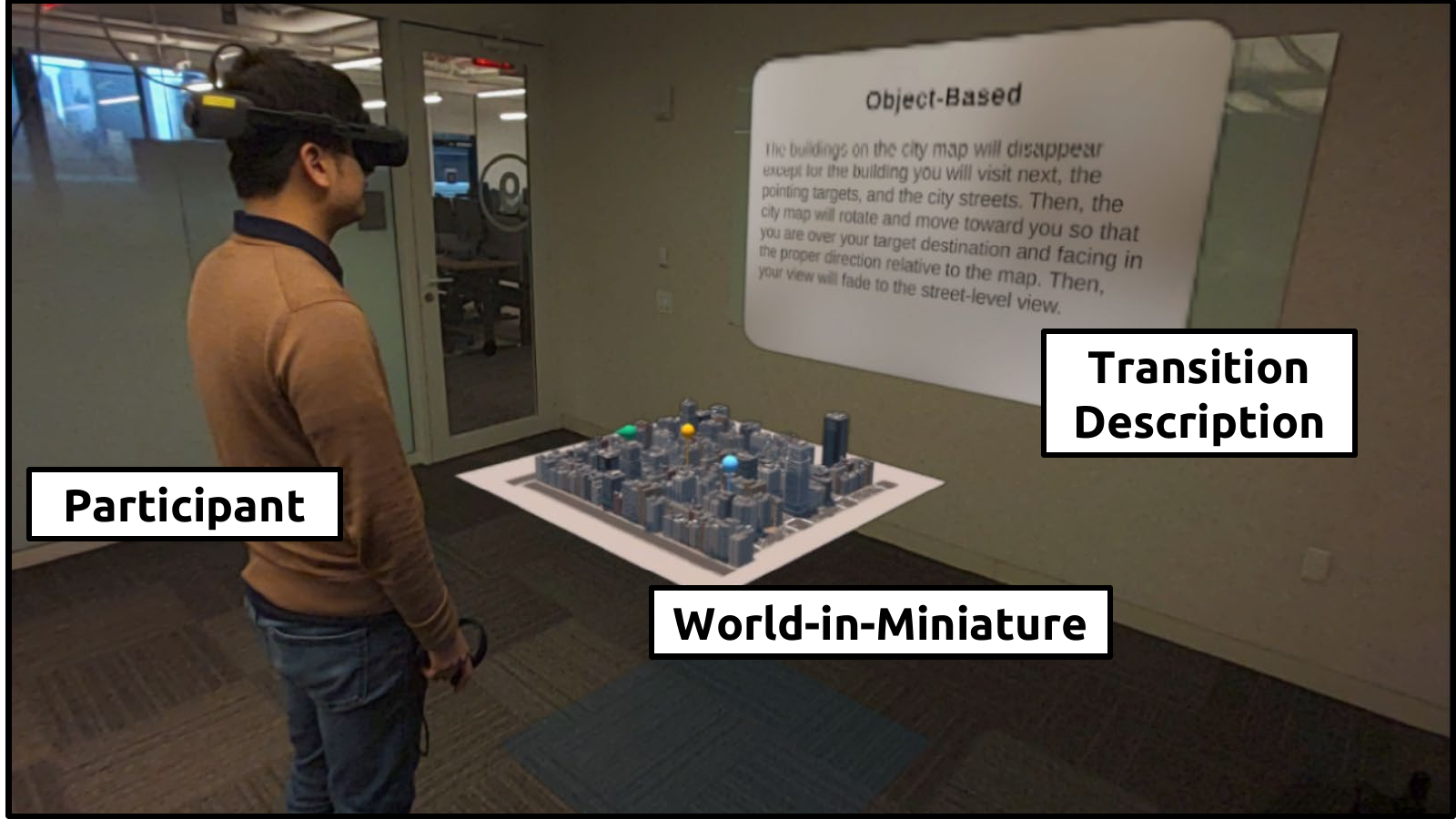}
    \caption{A participant at the beginning of a trial.}
    \label{fig:exp_setup}
    \cvspace{-1ex}
\end{figure}

\subsection{Experiment Environments and Viewpoints}
\paragraph{\textbf{\wimview{}}}
Each VE was an approx. $400\times400$ meter section of a virtual city environment available on the Unity asset store.\footnote{\url{https://assetstore.unity.com/packages/3d/environments/urban/japanese-city-240217}}
In the \wimview{}, which is how each trial began, the VE was shown at $1:400$ scale floating $1$\,m from the floor in the experiment space, and it included POI and landmark pins.

\paragraph{\textbf{POI Pin}}
POIs were marked with green pins.
Each consisted of a thin cylinder with a sphere on top, and a cone was attached to the sphere pointing toward the building that the user would be facing after they transitioned to the \streetview{} for the POI.
This was also the building that the user would enter after transitioning to the \insideview{}.
We applied different rotation degrees and directions to the POI pins within and across environments to add variety to the experiment and avoid participants becoming too accustomed to the kind of rotation they experienced when transitioning between scales.
Each POI pin was rotated either $\pm45\degree$, $\pm90\degree$, or $\pm135\degree$ with respect to the WiM (a negative angle indicates a leftward rotation).
Each VE had one POI with each rotation magnitude.
Two environments contained two POI pins rotated to the left and one pin rotated to the right, and vice versa for the other two environments.
The order in which a participant visited each POI in a VE was randomized.

\paragraph{\textbf{Landmark Pins}}
The \wimview{} of an environment included an orange and a blue landmark pin that were used in spatial orientation tests for each POI pin.
These pins were positioned in different locations relative to each POI pin to avoid participants becoming too familiar with the spatial recall task.
The positions for the landmark pins for a given trial were randomly and exclusively selected from a set of six candidate positions that were unique to each VE and set of POIs.
None of the candidate positions were on the same street as any of the POI pins in an environment to ensure that they would not be within participants' line-of-sight after they transitioned to a POI's \streetview{}.

\paragraph{\textbf{\streetview{}}}
After transitioning to the POI's \streetview{}, a wireframe shader was applied to the building in front of the POI.
The room that the participant would enter appeared as a cyan-colored cube in the building wireframe, and a smaller yellow cube with a purple arrow extending out from it pointed in the direction that the participant would be facing inside the building (serving the same purpose as the cone that indicated the direction from the \wimview{}).
We applied different orientation changes to participants' resulting direction inside the room in the same way as for the POI pins, i.e., each \insideview{} orientation was rotated either $\pm45\degree$, $\pm90\degree$, or $\pm135\degree$ with respect to the user's direction to the building from their \streetview{} position (one of each per VE, and left/right rotations balanced across all VEs).
For each POI, the same-scale rotation differed from the multi-scale rotation in both direction (reversed) and magnitude, ensuring participants did not experience identical consecutive rotations.

\paragraph{\textbf{\insideview{}}}
Each \insideview{} placed the participant inside an enclosed furnished office room \new{on an upper floor of the target building} containing common objects such as desks, cabinets, shelving, and potted plants, with two windowed walls offering partial visibility to the surrounding environment.
\new{Each same-scale transition thus involved vertical movement from street level, analogous to riding an elevator.}
Room layouts and furnishings varied across POIs and environments to prevent participants from using room recognition as an orientation cue.

\begin{table}[t!]
\caption{Steps of each trial.}
\label{tab:trial_procedure}
\centering
\small
\cvspace{-2ex}
\begin{tabular}{|l|l|l|}
\hline
\textbf{Step} & \textbf{Content / Activity} & \textbf{Description}                    \\ \hline
1    & View 1             & \wimview{}                     \\ \hline
2    & Transition 1       & Multi-Scale Transition         \\ \hline
3    & View 2             & \streetview{}                  \\ \hline
4    & Spatial Test 1     & Landmark Orientation Test      \\ \hline
5    & View 3             & \streetview{} with wireframe   \\ \hline
6    & Transition 2       & Same-Scale Transition          \\ \hline
7    & View 4             & \insideview{}                  \\ \hline
8    & Spatial Test 2     & Last Position Orientation Test \\ \hline
\end{tabular}
\cvspace{-3ex}
\end{table}

\subsection{Procedure}

After providing informed consent, participants were briefly introduced to the experiment's structure and task.
They then completed the computerized spatial orientation test~\cite{friedman2020computerized} on a laptop.
Next, the experimenter explained the controller inputs and interaction mechanics.
Participants then put on the XR HWD and completed a training trial in a separate simplified environment using the baseline \tFade{} transition.
The experimental trials then began following the steps outlined in~\secref{sec:userstudy} and~\tabref{tab:trial_procedure}.
After completing the three trials for each condition, participants completed the questionnaires in VR, yielding one set of responses per technique.
A semi-structured interview was conducted after all four conditions were completed (see the Supplementary Materials for specific questions).

\subsubsection{Spatial Orientation Tests}
\label{sec:spatialtest}

We included spatial recall tasks inspired by related work to assess participants' ability to orient themselves within the VE~\cite{shayman2024effects, peck2010ird, gottsacker2024residue, bhandari_teleportation_nodate}.
After each transition, a pop-up window prompted participants to point in the direction of the spatial recall target(s) relative to their new position and heading.
The indicated direction was recorded as the controller's forward vector at the moment of trigger press.
No feedback on accuracy was provided.
For multi-scale transitions from \wimview{} to \streetview{}, participants pointed to each of the two \textbf{landmark pins} in sequence.
For same-scale transitions from \streetview{} to \insideview{}, the target was the participant's \textbf{last virtual position}, i.e., the street-level position they had just transitioned from.

\subsubsection{Transition Implementations}

To isolate the effects of the transition techniques, we standardized all transition durations.
Each transition took $19\,s$, followed by a $1.5\,s$ fade-to-black; this duration was consistent across both multi-scale and same-scale transitions.
\new{All motion in transition animations used a cubic ease-in/ease-out velocity function to avoid abrupt visual motion.}
These timings were determined through pilot testing; different durations may be required to smoothly accomplish the transitions in different settings, e.g., across different scale differences, or when presenting supplemental narrative information.
With the \tFade{} baseline, participants were given the same $19\,s$ to study the environment and target viewpoint indicators before the fade-out began, ensuring that all conditions afforded equal time for spatial orientation regardless of whether a transition visualization was provided.
\new{We note that equal transition duration does not necessarily entail equal cognitive processing opportunity. The techniques differ in information density and temporal structure, so a fixed duration may suit some techniques better than others (see \secref{sec:limitations}).}

\subsection{Measures}

\subsubsection{Spatial Orientation Ability}

Participants completed the computerized spatial orientation test (SOT)~\cite{friedman2020computerized} prior to the experimental trials.
Higher scores indicate worse spatial orientation ability.
Scores were included as a covariate in the analysis to account for individual differences in baseline spatial orientation ability.

\subsubsection{Spatial Recall Accuracy}

We quantified spatial recall accuracy using absolute angular error on the spatial recall tests described in~\secref{sec:spatialtest}.
For each pointing response, we computed the unsigned yaw-axis angular difference between the indicated direction and the ground-truth direction to the target from the participant's position at the time of response, yielding values in the range $[0\degree, 180\degree]$.
For multi-scale transitions, the two landmark pin responses per trial were treated as separate observations, producing two angular error measurements per trial.

\subsubsection{Questionnaires}

\begin{itemize}[leftmargin=1.5em, nosep]
    \item \textbf{Discomfort:}
    We measured participant discomfort using a single-item Discomfort Scale adapted by Wu et al.~\cite{wu2021discomfort}: ``Please rate the discomfort level you are experiencing now on a level of 0 (no discomfort at all) to 10 (severe discomfort).''
    \item \textbf{User Experience:}
    The User Experience Questionnaire short version (UEQ-S)~\cite{schrepp2017ueqs} captured pragmatic quality (efficiency, perspicuity, dependability) and hedonic quality (originality, stimulation).
    \item \textbf{Workload:}
    The NASA Task Load Index (TLX)~\cite{hart2006tlx} assessed perceived workload across six dimensions: mental demand, physical demand, temporal demand, effort, performance, frustration.
    \item \textbf{Continuity:}
    A continuity questionnaire adapted from Husung and Langbehn~\cite{husung2019portals} evaluated participants' sense of an unbroken, coherent experience throughout the transition.
\end{itemize}

\section{Quantitative Results}
\label{sec:quantresults}

We analyzed spatial recall data using Bayesian mixed-effects models~\cite{burkner_brms_2017} and questionnaire data using Aligned Rank Transform~\cite{wobbrock2011art}.

\begin{figure}[t!]
    \centering
    \includegraphics[width=\linewidth]{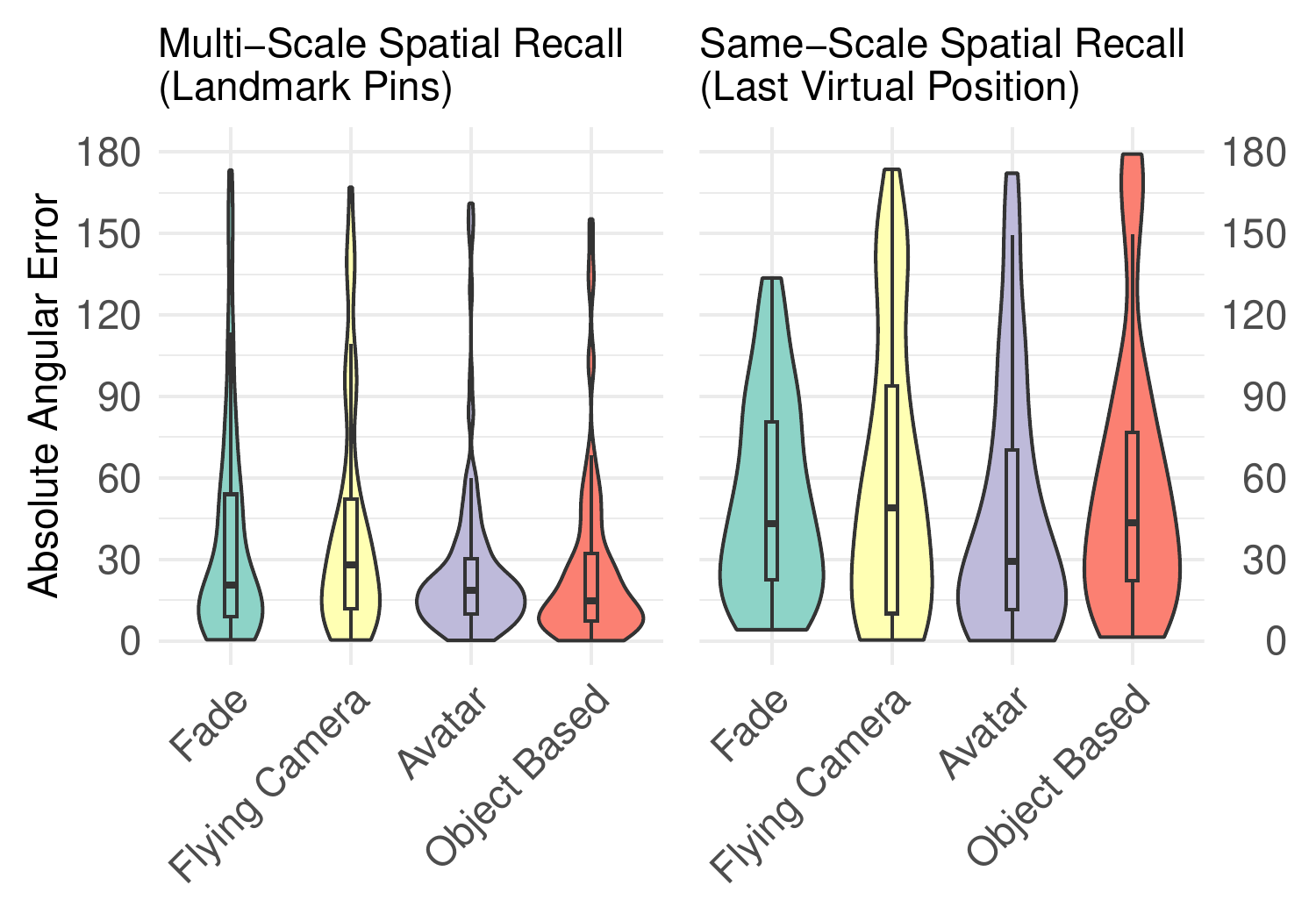}
    \cvspace{-4ex}
    \caption{Spatial recall error across transition types. (Left) Landmark pin spatial recall error after transitioning from the \wimview{} to the \streetview{}.
    (Right) Last virtual position recall error after transitioning from the \streetview{} to the \insideview{}.
    Violin plots show the distribution of absolute angular error.
    Overlaid box plots represent the interquartile range (IQR) (middle 50\%); solid line indicates median.
    }
    \cvspace{-1ex}
    \label{fig:angles_combined}
\end{figure}

\begin{figure*}[t!]
    \centering
    \includegraphics[width=0.84\linewidth]{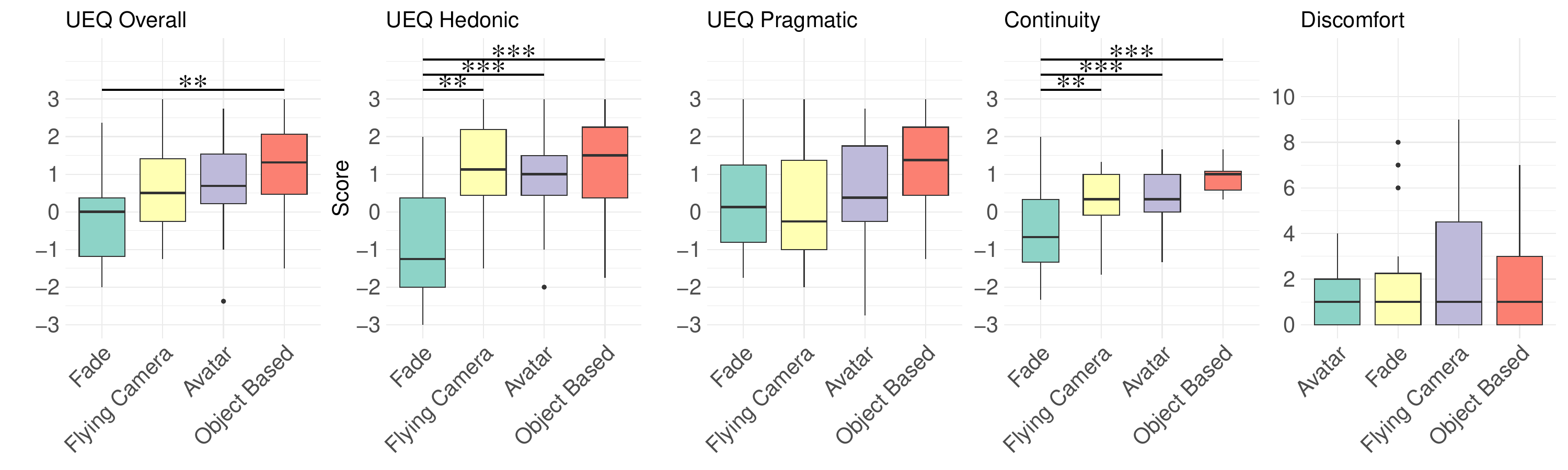}
    \cvspace{-2ex}
    \caption{Boxplots of subjective questionnaire results across techniques. Higher is better for all scales except Discomfort.
Significance markers denote pairwise comparisons ($^{*}p<.05$, $^{**}p<.01$, $^{***}p<.001$).
}
    \cvspace{-1ex}
    \label{fig:uxq}
\end{figure*}

\subsection{Spatial Recall Accuracy}

To analyze the effects of transition techniques on participants' angular error on the spatial memory tasks, we used Bayesian mixed-effects models~\cite{burkner_brms_2017}.
We chose a Gamma likelihood with a log link function to accommodate the strictly positive, right-skewed distribution of angular errors.
For both multi-scale and same-scale angular errors, we fit a Gamma regression model predicting absolute yaw error from Transition Type, Rotation Angle (the absolute rotation magnitude between pre- and post-transition heading, treated as a categorical factor with levels $45\degree$, $90\degree$, and $135\degree$), and SOT Score with a random intercept for participant:
\[
\text{YawError} \sim 
    \text{Technique}
    + \text{RotationAngle}
    + \text{SOT\_Score} 
    + (1|\text{PID})
\]
Each model was fit using four chains of 5,000 iterations each, half of which were warmup iterations.
We specified weakly informative priors to balance regularization and flexibility, reflecting plausible ranges based on the scale of the data and domain knowledge.
For each model, we used normal$(0,2.5)$ priors on the intercept and slopes on the \emph{log-mean} scale (a $95\%$ prior interval of approximately $\pm4.9$ implies very broad multiplicative effects on the mean response, about $\exp(\pm4.9)\approx 0.007$ to $134$), providing gentle regularization without restricting plausible effect sizes. We used an exponential$(1)$ prior on the random-intercept SD (also on the log scale), which shrinks toward zero yet allows substantial between-participant heterogeneity (95th percentile $\approx 3$), i.e., wide variation in participant baselines. We used a Gamma$(2,1)$ prior on the Gamma shape parameter which has mean $2$ (95\% prior mass roughly $0.24$–$5.57$), permitting a wide range of right-skew dispersion and ruling out implausible extremes.

We verified model adequacy through prior and posterior predictive checks, convergence diagnostics ($\hat{R}$, ESS, trace plots, divergent transitions), sensitivity analyses, and LOO cross-validation.
All model chains converged ($\hat{R}=1.00$) with large effective sample sizes.
Because the models use a log link, coefficients are expressed on the log scale but are most interpretable multiplicatively. For clarity, we report back-transformed results as approximate percentage changes in angular error ($100\cdot(\exp\beta-1)$).
Effects are considered credible when their 95\% credible interval (CrI) excludes zero; effects whose CrI includes zero are described as uncertain.
Spatial recall results are visualized in~\figref{fig:angles_combined}.

\subsubsection{Multi-Scale Transition Spatial Recall}

The model was fit on $N=480$ observations from $20$ participants. 
The estimated Gamma shape was $1.03$ ($95\%$ CrI $[0.92,\,1.15]$), consistent with right-skewed errors and close to an exponential distribution, supporting the distributional choice.

\textbf{Transition Type.} 
Posterior estimates indicated that \tAvatar{} ($M=26.0\degree$, $SD=29.0\degree$) and \tObj{} ($M=26.9\degree$, $SD=32.1\degree$) led to lower angular error than \tFade{} ($M=37.4\degree$, $SD=40.3\degree$) and \tFlycam{} ($M=41.3\degree$, $SD=41.9\degree$).  
Relative to \tFade{}, \tAvatar{} reduced error by about $27\%$ (95\% CrI $[-43\%,-7\%]$), and \tObj{} reduced error by about $30\%$ ($[-46\%,-10\%]$). \tFlycam{} was uncertain relative to \tFade{} (about $-4\%$, 95\% CrI $[-26\%,+24\%]$).  
Pairwise comparisons among the non-referent conditions showed that \tFlycam{} led to higher error than both \tAvatar{} ($+43\%$, 95\% CrI $[+11\%,+85\%]$) and \tObj{} ($+49\%$, 95\% CrI $[+15\%,+91\%]$). The contrast between \tAvatar{} and \tObj{} was small and uncertain ($+4\%$, 95\% CrI $[-19\%,+34\%]$).

\textbf{Rotation Angle and SOT.} 
Effects for Rotation Angle were uncertain ($90\degree$: about $-9\%$, 95\% CrI $[-27\%,\,+14\%]$; $135\degree$: about $+14\%$, 95\% CrI $[-9\%,\,+42\%]$, both relative to $45\degree$). SOT score showed a positive association of about $+1\%$ per point (95\% CrI $[0\%,\,+3\%]$), consistent with the expected direction given that higher SOT scores indicate poorer spatial ability.

\textbf{Random effects.} 
Participants varied in baseline error with a multiplicative standard deviation of about $+43\%$ (95\% CrI $[+23\%,+75\%]$).

\subsubsection{Same-Scale Transition Spatial Recall}

The model was fit on $N=228$ observations from $19$ participants (one excluded due to a logging error that corrupted their same-scale pointing data).
The estimated Gamma shape was $1.11$ ($95\%$ CrI $[0.93,\,1.31]$), consistent with right-skewed errors.

\textbf{Transition Type.}
Descriptive means were similar across conditions: \tAvatar{} ($M=47.8\degree$, $SD=47.4\degree$), \tFade{} ($M=51.5\degree$, $SD=36.5\degree$), \tObj{} ($M=56.7\degree$, $SD=49.4\degree$), and \tFlycam{} ($M=58.9\degree$, $SD=51.1\degree$).
Posterior estimates confirmed no credible differences among transition types.
Relative to \tFade{}, \tAvatar{} showed a small uncertain reduction ($-12\%$, 95\% CrI $[-38\%,+25\%]$), while \tFlycam{} ($+12\%$, $[-22\%,+58\%]$) and \tObj{} ($+11\%$, $[-22\%,+58\%]$) showed small uncertain increases.

\textbf{Rotation Angle and SOT.}
Larger rotation angles were associated with higher error: $90\degree$ showed an uncertain increase of about $+23\%$ (95\% CrI $[-10\%,+68\%]$), while $135\degree$ showed a credible increase of about $+49\%$ (95\% CrI $[+11\%,+103\%]$), both relative to $45\degree$.
SOT score showed no meaningful association ($0\%$ per point, 95\% CrI $[-1\%,+2\%]$).

\textbf{Random effects.}
Participants varied in baseline error with a multiplicative standard deviation of about $+39\%$ (95\% CrI $[+11\%,+77\%]$).

\subsection{Questionnaire Results}

We averaged questionnaire responses per Participant~$\times$~Technique and analyzed the data using the Aligned Rank Transform~\cite{wobbrock2011art} via the \texttt{ARTool} R package, with ParticipantID as a within-subject error term.
Where significant effects of Technique were found, we conducted post hoc pairwise comparisons using estimated marginal means with Holm correction.
Partial $\eta^2$ is reported for effect size.
Group means and standard deviations are reported on the original response scale.
Questionnaire results are visualized in~\figref{fig:uxq}.

\subsubsection{Workload (NASA-TLX)}
No significance was found for the mental, physical, performance, effort, and frustration sub-scales ($p>.05$).
Temporal demand showed a significant main effect of Technique, $F(3,57)=3.92$, $\mathbf{p=.013}$, $\eta_p^2=.17$. Post-hoc comparisons showed \tFade{} ($M=53.8$, $SD=23.9$) was rated significantly higher than \tObj{} ($M=39.4$, $SD=21.4$), $\mathbf{p=.018}$. No other contrasts were significant. 

\subsubsection{User Experience (UEQ)}

\paragraph{\textbf{Overall}}
There was a significant effect of Technique on Overall UEQ scores, $F(3,57)=5.56$, $\mathbf{p=.002}$, $\eta_p^2=.23$.
Descriptively, \tObj{} was rated highest ($M=1.24$, $SD=1.25$), followed by \tAvatar{} ($M=0.74$, $SD=1.18$) and \tFlycam{} ($M=0.72$, $SD=1.34$), with \tFade{} lowest ($M=-0.21$, $SD=1.17$).
Post hoc tests indicated \tObj{}$>$\tFade{} ($\mathbf{p=.001}$).
The comparison between \tAvatar{} and \tFade{} approached significance ($p=.054$), while the comparison between \tFlycam{} and \tFade{} was not significant ($p=.115$).
No other pairwise contrasts were significant.

\paragraph{\textbf{Pragmatic}}
There was no significant effect of Technique on UEQ Pragmatic ratings, $F(3,57)=2.49$, $p=.070$, $\eta_p^2=.12$.
Descriptively, \tObj{} was rated highest ($M=1.26$, $SD=1.28$), followed by \tAvatar{} ($M=0.59$, $SD=1.39$), \tFade{} ($M=0.25$, $SD=1.30$), and \tFlycam{} ($M=0.20$, $SD=1.60$).

\paragraph{\textbf{Hedonic}}
There was a significant main effect of Technique on UEQ Hedonic ratings, $F(3,57)=11.18$, $\mathbf{p<.001}$, $\eta_p^2=.37$.
All three techniques were rated higher than \tFade{}: \tObj{}$>$\tFade{} ($\mathbf{p<.0001}$), \tFlycam{}$>$\tFade{} ($\mathbf{p<.0001}$), and \tAvatar{}$>$\tFade{} ($\mathbf{p=.001}$).
Descriptively, \tFlycam{} ($M=1.24$, $SD=1.28$) and \tObj{} ($M=1.21$, $SD=1.48$) were rated similarly, followed by \tAvatar{} ($M=0.89$, $SD=1.24$), with \tFade{} lowest ($M=-0.68$, $SD=1.58$).
No pairwise contrasts among the three techniques were significant.

\subsubsection{Continuity}
There was a significant main effect of Technique on Continuity scores, $F(3,57)=8.23$, $\mathbf{p<.001}$, $\eta_p^2=.30$.
Descriptively, \tObj{} had the highest continuity scores ($M=0.82$, $SD=0.75$), followed by \tAvatar{} ($M=0.32$, $SD=0.92$) and \tFlycam{} ($M=0.30$, $SD=0.80$), with \tFade{} rated lowest ($M=-0.52$, $SD=1.14$).
Pairwise comparisons indicated that all three techniques were rated higher than \tFade{}: \tObj{}$>$\tFade{} ($\mathbf{p<.0001}$), \tFlycam{}$>$\tFade{} ($\mathbf{p=.036}$), and \tAvatar{}$>$\tFade{} ($\mathbf{p=.036}$).
Comparisons of \tObj{} with \tFlycam{} ($p=.087$) and with \tAvatar{} ($p=.087$) did not reach significance.
The contrast between \tFlycam{} and \tAvatar{} was negligible ($p=.944$).

\subsubsection{Discomfort}
There was no significant main effect of Technique on Discomfort scores, $F(3,57)=1.47$, $p=.232$, $\eta_p^2=.07$.
Mean discomfort was lowest for \tAvatar{} ($M=1.15$, $SD=1.23$), followed by \tFade{} ($M=1.85$, $SD=2.46$) and \tObj{} ($M=1.85$, $SD=2.28$), and was highest for \tFlycam{} ($M=2.55$, $SD=2.98$).

\section{Qualitative Results}
\label{sec:qualresults}

We analyzed the semi-structured interview transcriptions using a hybrid deductive-inductive thematic approach~\cite{fereday_muircochrane_2006_thematic} grounded in the a priori domains of Orientation Strategies, Transition Preferences, and Discomfort.
Within each domain, participant statements were iteratively reviewed and synthesized to identify recurring patterns, shared strategies, and points of divergence across participants.
Related statements were grouped into subthemes reflecting specific strategies, experiences, and cognitive or physical effects.
Two researchers conducted this analysis independently and met to reconcile codes and subthemes, resolving disagreements through discussion.

\subsection{Orientation Strategies}

\paragraph{\textbf{Computational and spatial thinking strategies.}}
Participants employed diverse computational and spatial thinking strategies to maintain orientation during view transitions. The most prevalent approach involved angle- and rotation-based reasoning (N=4): participants computed relative rotations by remembering how much they turned, comparing positions by relative angle, and using structured heuristics such as clock-face positions or degree-based calculations. Some (N=2) used hybrid methods combining clock reasoning with color coding or cardinal directions using global reference, though one participant commented that finding north was not always helpful and sometimes increased task difficulty. Map and landmark-based strategies were also common (N=3), with participants anchoring orientation through building positions, block offsets, street context relative to arrow direction, and bird's-eye mental models. They noted orientation was easier at street level than inside buildings and viewing the map before transitions provided familiarity and comfort.

\paragraph{\textbf{Embodied and emergent sense-making}}

Beyond analytical strategies, participants engaged in embodied sense-making processes. Many described constructing strategies on-the-fly through conscious, deliberate effort (N=5). They paired their left/right hands with color-coded pins, pictured themselves oriented as the arrow to determine hand positions, or attempted to memorize physical hand locations when imagining pointing toward landmarks. However, these body-based mnemonics were not always reliable, with one participant reporting rapid decay of hand position memory after transitioning. Some participants also used transition cues to anticipate future facing direction and mentally or physically pre-orient (N=2).

\subsection{Transition Preferences}

\paragraph{\textbf{\tAvatar{}: Pose-based orientation}}

Participants used the avatar's body as a reference for orientation, including shoulder positioning and hand alignment with directional pointers to indicate target facing direction and relative pin placement (N=4). The avatar helped visualize the transition (N=2). However, two found it only weakly helpful, preferring a persistent ``ghost'' of the prior location, or reported not feeling connected to it, describing it as disembodied with potentially disorienting extra visuals. 
One noted that the avatar moved too quickly before they fully processed the transition.

\paragraph{\textbf{\tObj{}: Reference cues and spatial context}}

The \tObj{} transition was effective for remembering reference points and positions, with the reduced map visualization particularly valued. Participants reported that objects provided clearer directional reference than avatars, and the ``skeleton mode'' showing interior positioning felt seamless. The map-first orientation followed by zooming was intuitive, and this method allowed participants to skip mental processing by directly showing the destination view. Several (N=3) characterized these cues as easier and smoother. One noted simultaneous visual changes introduced complexity in the same-scale transition.

\paragraph{\textbf{\tFlycam{}: Continuous spatial information}}

Participants were divided on \tFlycam{}. Some (N=2) found it distracting or over-engineered, feeling like ``looking at another screen within VR'' with limited grid visibility reducing utility. One reported it provided no more support than fade to black. Conversely, two others valued the fly-in experience as video-game-like, with the continuous spatial information providing more helpful orientation cues than discrete transitions.

\paragraph{\textbf{\tFade{}: Minimal support with memory carryover}}

Participants reported mixed preferences for \tFade{}. Several found it difficult, disconnected, or lacking continuity, increasing cognitive load without useful support. Others (N=2) perceived it as easiest because it minimized distraction and unnecessary graphics. A key mechanism was memory carryover: the map image persisted in memory and supported orientation after the fade, though this lacked an experiential sense of travel and relied on mental reconstruction.

\subsection{Discomfort}

\paragraph{\textbf{Limited reports of discomfort}}

Most participants reported no motion sickness with the transitions (N=15).
\tAvatar{} and fade-to-black transitions were described as comfortable due to their resemblance to familiar screen changes.
Participants described discomfort emerging with prolonged use or particular visual properties (N=3). Visual discomfort was due to graphical shimmering or a ``buildup'' effect. 
\tFlycam{} was associated with motion sickness for two participants: one described the camera fly as potentially sickness-inducing and difficult/disorienting as well as distracting. Another reported discomfort with flying or elevator-like movement.

\section{Discussion}

\subsection{Summary of Findings}

For multi-scale spatial recall, \tAvatar{} and \tObj{} reduced angular error relative to \tFade{}, while \tFlycam{} showed no credible difference from \tFade{} and produced higher error than both \tAvatar{} and \tObj{}.
This partially extends the findings of Rahimi et al.~\cite{rahimi_scene_2020}, who showed that animated interpolation supported better spatial awareness than teleportation in system-automated transitions. However, \tFlycam{} (the technique most analogous to animated interpolation) did not improve spatial recall, while \tAvatar{} and \tObj{} did, suggesting that destination previewing \new{was} a more effective mechanism for orientation than virtual travel animations in \new{our fixed-duration} multi-scale externally-guided context.
For same-scale spatial recall (i.e., \streetview{} to \insideview{}), transition technique did not reliably affect angular error.

For user experience, \tObj{} improved overall UEQ scores relative to \tFade{}, all three techniques improved hedonic ratings relative to \tFade{}, and continuity ratings were highest for \tObj{}, with \tAvatar{} and \tFlycam{} also exceeding \tFade{}.
This is consistent with prior findings that predictable transitions with destination previews improve continuity~\cite{husung2019portals} and that prominent transitions are particularly valued when users are unfamiliar with the target environment~\cite{pointecker_bridging_2022}.
Workload and comfort differences were largely absent: only temporal demand differed, with \tFade{} rated higher than \tObj{}, and discomfort showed no significant effect of transition technique, though \tFlycam{} exhibited the highest mean discomfort.

These findings provide partial validation of our design considerations.
The spatial orientation benefits of \tAvatar{} and \tObj{} support the value of maintaining spatial orientation through externalized reference frames and gradual scene introduction (DC-1).
Discomfort scores did not differ reliably across techniques, suggesting that none of the three designs introduced substantial visually-induced conflict (DC-2), though qualitative reports indicate that \tFlycam{} was less comfortable for some participants.
The hedonic and continuity advantages of the non-fade techniques over \tFade{} confirm that users perceive and value transition aesthetics (DC-3).
Finally, the superior spatial recall performance of \tAvatar{} and \tObj{} (which provide explicit destination previews) supports the utility of feedforward cues in externally-guided contexts where users cannot anticipate the transition outcome (DC-4).

\subsection{Design Implications}
\label{sec:implications}

The following implications are grounded in our empirical findings and are intended to guide the design of viewpoint transitions in externally-guided XR experiences.

\subsubsection{Transition Benefits for Large Spatial Transformation}

\new{Multi-scale transitions bring users through changes in position, heading, and scale. Our transition visualizations provided orientation benefits for this case.
The absence of technique differences for same-scale transitions may be specific to the arrangement we tested (\secref{sec:limitations}).}
\textbf{Takeaway: \new{Orientation benefits of transition techniques emerged for multi-scale transitions but not in our same-scale arrangement. In contexts involving large spatial transformations, such as crossing scales or entering unfamiliar environments, transition visualizations are most worthwhile, while simpler transitions may suffice for smaller viewpoint changes.}}

\subsubsection{Destination Previews Over Path Visualization}

\new{\tObj{} and \tAvatar{} kept the user's viewpoint stationary while previewing the destination viewpoint, whereas \tFlycam{} moved a virtual camera along the travel path to show the route.
Because users could not control the camera's movement, \tFlycam{}'s worse orientation scores may partly reflect the fixed 19\,s presentation tempo rather than an inherent weakness of virtual travel visualization, and longer or user-controlled transitions could alter this relationship.
Our findings extend destination previewing (previously shown to support orientation in self-guided teleportation and scale transitions~\cite{griffin_out--body_2019, huang_preview_2025, weissker_try_2024}) to externally-guided multi-scale contexts.
Additionally, Lee et al.~\cite{lee_designing_2023} found that separating rotation and translation phases during scale transitions is critical for maintaining spatial orientation. \tObj{} and \tAvatar{} visualize rotation before translation, whereas \tFlycam{} at times interpolates both concurrently, which may partly explain the spatial recall differences.}
\textbf{\new{Takeaway: For fixed-duration externally-guided multi-scale transitions, \tObj{} and \tAvatar{}, which kept the viewpoint stable and previewed the destination state (position, heading, and surroundings), supported stronger orientation than \tFlycam{}'s dynamic travel visualization. In contexts where the audience's orientation is most important, destination-preview techniques are preferable.}}

\subsubsection{Aesthetic Benefits of Transitions}

All three techniques outperformed \tFade{} on hedonic UEQ and continuity ratings, including \tFlycam{} despite its weaker spatial recall performance.
This indicates that even transitions that do not improve orientation can still improve the felt quality of the experience.
\textbf{Takeaway: Spatially continuous visualizations provided meaningful hedonic and continuity benefits over abrupt cuts. In contexts where engagement and perceived polish are primary goals (e.g., entertainment, tourism, or narrative XR) coherent transition visualizations provide experiential benefits over a fade-to-black.}

\subsubsection{Low Comfort Risk for Stationary Users}

Discomfort did not differ significantly across techniques, but \tFlycam{} exhibited the highest mean discomfort and drew two qualitative reports of motion sickness.
Externalizing motion to a decoupled camera feed does not fully eliminate vection risk. Some users still experience discomfort watching camera motion, particularly vertical or elevator-like movement.
For comfort-sensitive populations or prolonged use, stationary-user techniques such as \tObj{} and \tAvatar{} may carry lower risk.
\textbf{Takeaway: Techniques that keep the user's viewpoint stationary produce less discomfort than those involving observed camera motion, even when the motion is decoupled from the user's perspective.}

\subsection{Limitations}
\label{sec:limitations}

Position, orientation, and scale covaried across transitions in our study scenario, so we cannot isolate the contribution of each dimension to the observed technique differences.
Future work could systematically vary these dimensions independently to identify which spatial changes most benefit from transition support.
\new{The same-scale transitions always involved elevator-like vertical movement into an enclosed upper-floor room, which may introduce additional orientation demands that differ from other same-scale scenarios. Additionally,} the \insideview{} rooms contained windows through which some exterior geometry was visible, which could have supplemented recalled orientation with visual inference.
Because room orientations differed across POIs and conditions, this cue was not consistently informative, but it may have reduced sensitivity to technique differences on the same-scale spatial recall task.
The fixed 19\,s transition duration was applied uniformly across techniques and scale conditions, meaning that the rate of visual change varied; techniques or transitions involving more spatial distance or visual complexity may have felt rushed, while simpler ones may have afforded more processing time.
The sample of 20 participants and the use of single-item measures may have limited sensitivity to smaller effects, particularly for workload and comfort, even with the within-subjects design.
Finally, the study \new{held several scenario parameters constant:} environment type (urban), scale ratio (1:400 WiM to full scale)\new{, progression (overview to street to indoor), pacing (19\,s experimenter-triggered transitions), and audience configuration (a single passive user).
This scenario is representative of common XR presentation use cases such as architectural walkthroughs, urban planning reviews, and lecture-style overviews, and we expect the findings to transfer best to similar contexts.
Future work could test generalizability to other scenarios.}

\subsection{Future Work}

We see promising directions for future related research.
Our qualitative findings revealed diverse orientation strategies (angle-based reasoning, embodied mnemonics, map-based anchoring), suggesting that no single transition design is optimal for all users. Future transitions could adapt cue sets to users' inferred strategies. Adaptive interfaces such as interaction adaptation~\cite{lai_adaptique_2025} and environment summarization~\cite{gunturu_realitysummary_2025} may support this direction as information spaces grow in complexity.
Additionally, broader evaluations should test multi-scale transitions under higher workload, collaborative or interactive presentation scenarios, and longer durations to capture learning effects and cumulative discomfort\new{, as well as reversed (e.g., indoor-to-overview) or randomized progression orders}.
\new{Investigating alternative multi-scale transition designs and other dimensions of Bowman et al.'s travel taxonomy~\cite{bowman1997travel} in externally-guided contexts, such as different velocity profiles, is another promising direction.}
Finally, there are open questions around audience integration. For instance, future transitions could provide limited audience control (e.g., pacing adjustments) to accommodate individual comfort thresholds and strategy formation time.

\section{Conclusion}
We investigated how transition techniques affect spatial understanding, user experience, and comfort during externally-guided multi-scale viewpoint changes in which users lack control over the transition process.
Our evaluation of three techniques against a fade-to-black baseline found that transition design substantially affects spatial recall when users experience large changes in scale and viewpoint, but not in our same-scale arrangement.
The techniques that improved multi-scale recall (\tObj{} and \tAvatar{}) externalized the mental rotation and pose alignment that users would otherwise need to perform, providing stable reference frames and explicit destination previews.
In contrast, \tFlycam{}'s virtual travel visualization did not provide orientation benefits.
Beyond spatial recall, all non-fade techniques improved hedonic and continuity ratings, indicating that transitions can improve experience even without orientation benefits.
While grounded in a specific scenario (presenter-triggered transitions across a city-scale WiM, street-level, and indoor view), we hope the principles of reference-frame stability, destination previewing, and stationary-viewpoint design are applicable to externally-guided XR contexts more broadly.


\bibliographystyle{abbrv-doi-hyperref}

\bibliography{00_references}

\section*{Disclaimer}
This paper was prepared for informational purposes by the Global Technology Applied Research center of JPMorgan Chase \& Co. This paper is not a product of the Research Department of JPMorgan Chase \& Co. or its affiliates. Neither JPMorgan Chase \& Co. nor any of its affiliates makes any explicit or implied representation or warranty and none of them accept any liability in connection with this paper, including, without limitation, with respect to the completeness, accuracy, or reliability of the information contained herein and the potential legal, compliance, tax, or accounting effects thereof. This document is not intended as investment research or investment advice, or as a recommendation, offer, or solicitation for the purchase or sale of any security, financial instrument, financial product or service, or to be used in any way for evaluating the merits of participating in any transaction.

\end{document}